\pdfoutput=1
\documentclass{article}

\PassOptionsToPackage{table}{xcolor}

\usepackage[final]{colm2026_conference}

\usepackage[T1]{fontenc}
\usepackage[utf8]{inputenc}
\usepackage{microtype}
\usepackage{graphicx}

\usepackage{booktabs}
\usepackage{tabularx}
\usepackage{multirow}
\usepackage{subcaption}
\usepackage{xcolor}
\usepackage{enumitem}
\usepackage{caption}
\usepackage{xurl} %
\usepackage{fontawesome5}
\usepackage{hyperref}
\hypersetup{hidelinks}

\newcommand{\dthrow}[1]{\textbf{#1}}

\definecolor{d2hInk}{HTML}{0A1424} %
\definecolor{d2hAccent}{HTML}{0055E9} %
\definecolor{d2hTint}{HTML}{A9CDFF} %

\newcommand{\dtwohhead}{%
  \footnotesize
  Doc2Harness: Benchmarking LLMs on Long-Horizon Harness Engineering of Coding Agents}
\makeatletter
\def\section{\@startsection{section}{1}{\z@}{-2.0ex plus -0.5ex minus -.2ex}%
  {1.5ex plus 0.3ex minus0.2ex}{\large\bf\raggedright\color{d2hAccent}}}
\def\subsection{\@startsection{subsection}{2}{\z@}{-1.8ex plus -0.5ex minus -.2ex}%
  {0.8ex plus .2ex}{\normalsize\bf\raggedright\color{d2hAccent}}}
\def\subsubsection{\@startsection{subsubsection}{3}{\z@}{-1.5ex plus -0.5ex minus -.2ex}%
  {0.5ex plus .2ex}{\normalsize\bf\itshape\raggedright\color{d2hAccent}}}
\makeatother
\renewcommand{\dthrow}[1]{\textbf{\color{d2hAccent}#1}}
\title{%
  \texorpdfstring{\raggedright}{}%
  \texorpdfstring{%
    \raisebox{-0.26\height}{\includegraphics[height=1.30em]{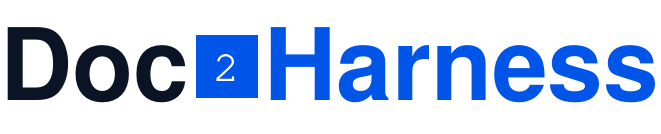}}%
  }{Doc2Harness}%
  : Benchmarking LLMs on Long-Horizon \mbox{Harness} Engineering of Coding Agents}

\author{%
  \textbf{Yuyao Ge}\textsuperscript{\normalfont 1},
  \textbf{Shenghua Liu}\textsuperscript{\normalfont 1,\textdagger},
  \textbf{Peizhong Gao}\textsuperscript{\normalfont 2},
  \textbf{Chendi Ge}\textsuperscript{\normalfont 2},
  \textbf{Chenxu Liu}\textsuperscript{\normalfont 3},
\\
  \textbf{Yiwei Wang}\textsuperscript{\normalfont 4},
  \textbf{Lingrui Mei}\textsuperscript{\normalfont 1},
  \textbf{Baolong Bi}\textsuperscript{\normalfont 1},
  \textbf{Jiayu Yao}\textsuperscript{\normalfont 1},
  \textbf{Jiafeng Guo}\textsuperscript{\normalfont 1},
  \textbf{Xueqi Cheng}\textsuperscript{\normalfont 1}
\\[0.9em]
  \textsuperscript{\normalfont 1}Institute of Computing Technology, Chinese Academy of Sciences \\
  \textsuperscript{\normalfont 2}Tsinghua University \quad \textsuperscript{\normalfont 3}Peking University \quad \textsuperscript{\normalfont 4}Southeast University
\\[0.6em]
  {\large\color{d2hAccent}\faGlobe\enspace\href{https://geyuyao.com/doc2harness/}{\textbf{geyuyao.com/doc2harness}}}}

\begin{document}

\makeatletter
\let\colm@titleblock\@maketitle
\def\@maketitle{\colm@titleblock \vspace{-0.30in}\begin{center}
\includegraphics[width=\textwidth]{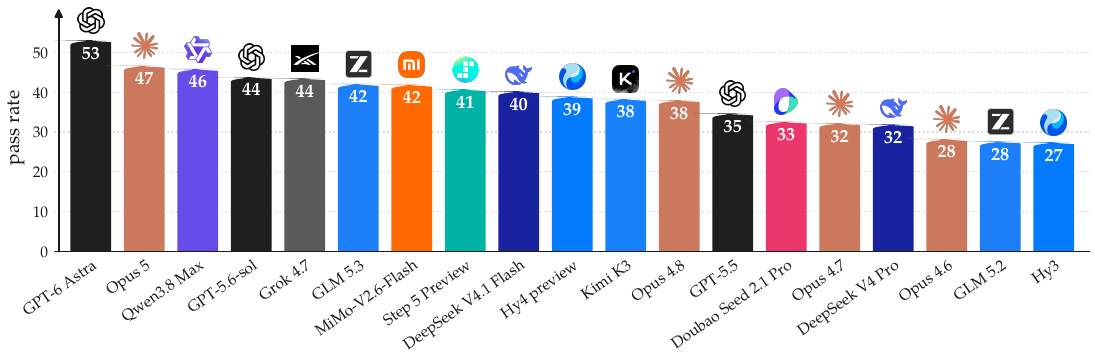}
\captionof{figure}{Pass rate of the nineteen models on the Doc2Harness benchmark, ordered from the strongest model on the left to the weakest on the right. Each model is scored on three independently built candidate harnesses, and each bar gives their mean.}
\label{fig:ranking}
\end{center}
}
\makeatother

\maketitle

\lhead{\dtwohhead}

{\renewcommand\thefootnote{}\footnotetext{\textsuperscript{\textdagger}\,Corresponding author.}}

\begin{abstract}
Agents driven by large language models are increasingly deployed on long-horizon tasks, where they use tools and modify real workspaces. How well they perform depends as much on the harness around the model as on the model itself. Existing benchmarks measure downstream task success under a fixed harness, leaving harness engineering itself unexplored. We propose Doc2Harness, a two-stage benchmark that evaluates long-horizon harness engineering directly. In Stage~1, a build model operating inside a fixed builder harness receives an interface-only contract skeleton and a development document that consolidates behavior across twelve frontier harnesses, and then constructs a complete candidate harness in an isolated container. In Stage~2, the frozen candidate harness runs in containers separate from the construction environment, one per test case, and is driven end to end by a scripted mock model. We score each test case using four observable channels, namely the requests sent to the model, the files on disk, the process exit code, and the command output; positive controls rule out vacuous passes. Across nineteen frontier models, each evaluated with three independent builds on 328 test cases, GPT-6 Astra achieves the highest pass rate at 53.05\%, while the weakest model reaches 27.44\%, a gap of roughly twenty-six percentage points. Even the strongest model fails nearly half of the test cases, and failures cluster around how cross-cutting functionality is split across modules. To our knowledge, Doc2Harness provides the first deterministic benchmark for measuring long-horizon harness engineering.
\end{abstract}

\section{Introduction}
\label{sec:introduction}

As coding agents driven by large language models (LLMs) become more capable, they are deployed in increasingly complex settings \citep{ge2025survey}. The harness therefore matters more: it manages the interaction loop, tool execution, context, permissions, session state, failure recovery, and result verification, and so determines how much of the model's capability reaches the environment \citep{mei2025contextsurvey}. At the same time, human supervision of every agent step has become difficult. This has motivated recursive self-improvement, in which an agent modifies the infrastructure that governs its own behavior. An agent executes a task, observes the outcome, modifies its own harness, and then uses the modified harness on the next task, improving itself over time. A model's ability to generalize this process may ultimately determine the capability of the resulting agent.

Building a harness is a type of software engineering (SWE) task, but it tests a kind of generalization that ordinary SWE tasks do not capture. At the data-source level, the open-source repositories used by SWE benchmarks are mostly mature projects that may have appeared repeatedly in training corpora. Harnesses, by contrast, have appeared only in the past few years and are updated at high frequency, so the resulting evaluation places greater emphasis on generalization. On the evaluation side, existing SWE benchmarks such as SWE-bench Pro \citep{swebenchpro}, NL2Repo-Bench \citep{nl2repobench}, and Commit0 \citep{commit0} judge whether the model's output is acceptable using the interactive unit tests supplied with the task. Such criteria do not transfer directly to harnesses, for two reasons.
(1) A single feature can be realized in many ways inside a harness, whereas an ordinary unit test observes behavior through a fixed interface.
(2) A harness does not complete a downstream task on its own, so its correctness must be assessed by driving it with a model.

We therefore propose Doc2Harness, a two-stage benchmark for evaluating how well LLMs develop harnesses. In Stage~1, the build model under evaluation operates inside a fixed builder harness and, given a development document and a contract skeleton that declares interfaces only, constructs a complete candidate harness inside an isolated container. In Stage~2, the candidate harness is moved into isolated containers, one per test case, launched and executed end to end as a black box, with a mock model providing responses according to a preset script and each test case running the harness through one complete session. We observe four channels for each test case, namely the requests sent to the model, the files on disk, the process exit code, and the command output, and aggregate the resulting statuses into a pass rate. The build model's ability to develop a harness is reflected indirectly in the pass rate.

Both inputs are designed to reduce reliance on parametric knowledge that the build model may already possess: the development document is rewritten from the user documentation of twelve frontier harnesses after merging and disambiguation, and the contract skeleton is distilled further from mini-swe-agent \citep{minisweagent} with all internal implementation removed. For Stage~2, we design several hundred test cases covering a wide range of scenarios. A test case may pass even though the feature under test never ran, a situation we call a vacuous pass; to guard against it we introduce positive controls that verify the observation preconditions. If a positive control fails, the observation precondition is not satisfied, and the test case is recorded as FAIL regardless of its own status.

\looseness=-1 %
As shown in Figure~\ref{fig:ranking}, we evaluate nineteen frontier models on Doc2Harness, scoring three candidate harnesses per model and ranking them by pass rate. The strongest, GPT-6 Astra, reaches a pass rate of 53.05\%, while the weakest reaches 27.44\%; this roughly twenty-six-point gap shows that the suite is neither saturated nor unable to discriminate among the models in this cohort. Across categories, a model's rank varies by as many as sixteen positions.

Our contributions are as follows.

\begin{itemize}
\item We propose Doc2Harness, which runs the build model under evaluation inside a fixed builder harness, gives it a development document and a contract skeleton that declares interfaces only, asks it to construct a complete candidate harness inside an isolated container, and then evaluates the result with several hundred test cases.
\item We provide a test suite that exercises a harness as a black box: the candidate harness is driven by a mock model following a preset script, and positive controls help rule out vacuous passes, keeping the evaluation accurate.
\item We evaluate nineteen frontier models and parse the source of every scored candidate harness. Higher-scoring candidate harnesses tend to place cross-cutting functionality in fewer modules. A model's rank also moves by up to sixteen positions across the six main categories.
\end{itemize}

\section{Doc2Harness}
\label{sec:benchmark}

\subsection{Task Description}
\label{sec:task}

\begin{figure}[t]
\centering
\includegraphics[width=\textwidth]{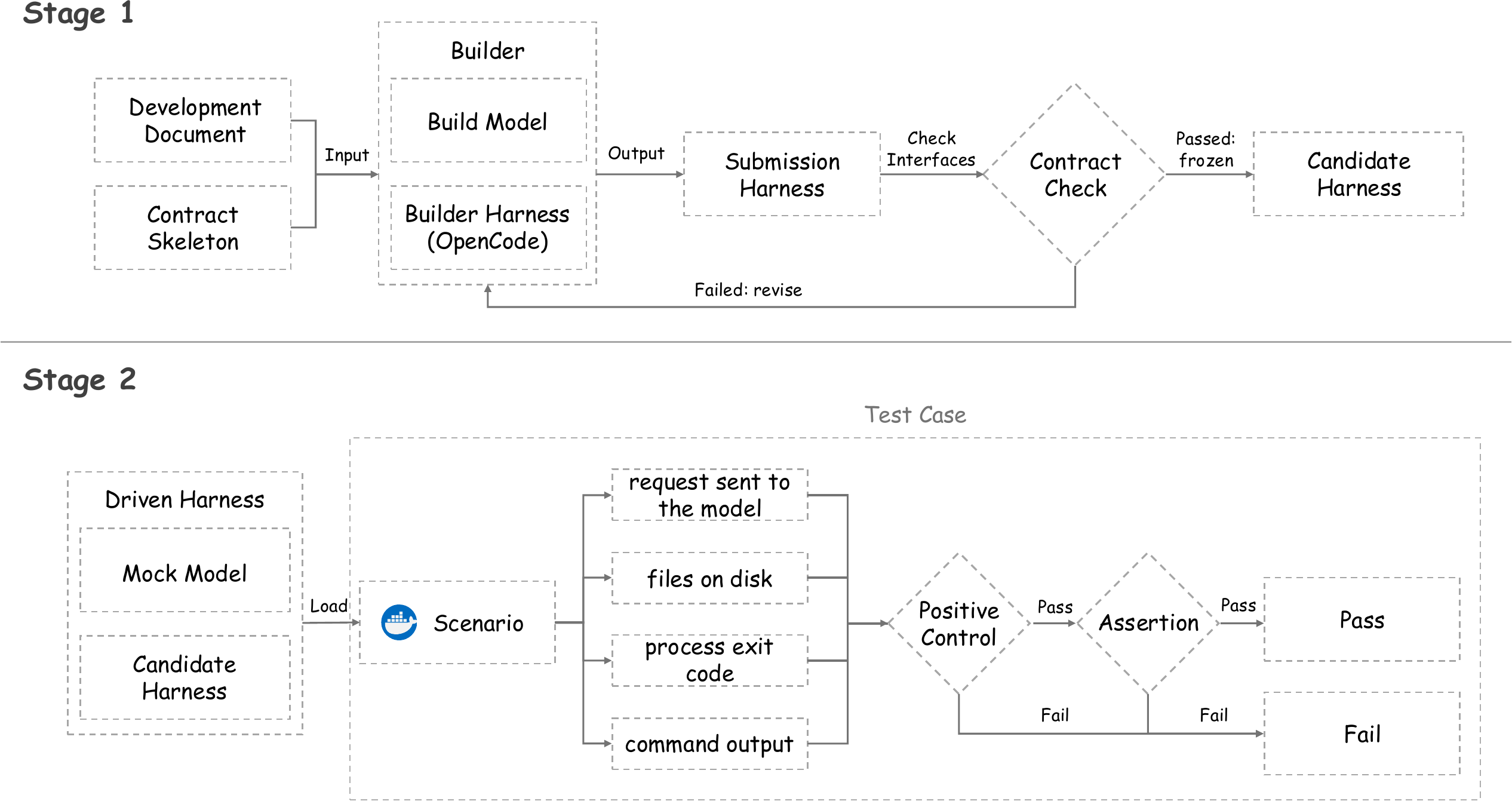}
\caption{Overview of Doc2Harness. Top (Stage~1): the builder, formed from the build model under evaluation and a fixed builder harness (OpenCode), takes the development document and the contract skeleton as input and emits a submission; the contract check verifies form rather than behavior, freezing the deliverable as the candidate harness on success and returning it to the builder on failure. Bottom (Stage~2): the frozen candidate harness and a mock model form the driven harness, which runs the scenario of each test case; effects are read from four channels, namely the requests sent to the model, the files on disk, the process exit code, and the command output, and the positive control gates the status before the assertion decides PASS or FAIL.}
\label{fig:pipeline}
\end{figure}

Doc2Harness has two stages, shown in Figure~\ref{fig:pipeline}.

\looseness=-1 %
\dthrow{Stage~1, Construction.} The \textbf{builder} combines the \textbf{build model} under evaluation with a fixed \textbf{builder harness}. It reads a \textbf{development document} and, starting from a \textbf{contract skeleton}, constructs a \textbf{candidate harness}. The development document prescribes the form and functionality the candidate harness should possess; the contract skeleton is a framework, simplified and adapted from mini-swe-agent \citep{minisweagent}, that retains only the interfaces and contains no implementation. The builder determines when Stage~1 ends by issuing a \textbf{submission}, without which the build does not count as delivered. At each submission, the contract check verifies that the deliverable compiles and installs, that all required commands are present, that no placeholder stubs remain, and that the three contract interfaces are in place, without running any test case. If the check passes, the build ends and the deliverable is frozen as the candidate harness, never to be modified again; if it fails, the deliverable is returned, and the builder may keep revising and submit again within its quota (\S\ref{sec:setup}). Each scored candidate harness represents the state the model submitted as finished, rather than a state reached only when the time limit expired.

\dthrow{Stage~2, Evaluation.} Inside containers isolated from the construction environment, one per test case, a \textbf{mock model} drives the candidate harness, and we determine the status of each test case in turn. Some test cases are additionally equipped with \textbf{positive controls}, which test whether an apparent pass reflects the required behavior rather than a failure of the observation mechanism (\S\ref{sec:pc}). A test case is recorded as PASS only when it passes on its own and every positive control gating it also passes. The build model's ability to develop a harness is reflected indirectly through the pass rate over test cases. We therefore rank build models rather than candidate harnesses.

Each stage holds one component fixed to reduce confounding. In Stage~1, the builder harness is uniformly the open-source coding agent framework OpenCode \citep{opencode} rather than harnesses such as Claude Code \citep{claudecode} or Codex CLI \citep{codexcli}, because such harnesses are developed by model vendors alongside their own models and may confer an advantage on the Claude and GPT model families. In Stage~2, the mock model supplies responses according to a preset script, which avoids interference from the instability that a real LLM introduces.

\subsection{Task Inputs}
\label{sec:inputs}

\subsubsection{Development Document}
\label{sec:doc}

The development document is written in three steps, namely collecting a feature inventory, turning that inventory into clauses, and disambiguating so that the clauses are self-consistent. The clause is the document's basic unit. Most clauses specify the external behavior of one feature, while the consistency rules specify relations across features and chapters.

\dthrow{Collecting a feature inventory.} In the first phase we crawl the user documentation of twelve frontier harnesses (Claude Code \citep{claudecode}, Kimi Code \citep{kimicode}, Gemini CLI \citep{geminicli}, Qwen Code \citep{qwencode}, OpenCode \citep{opencode}, OpenClaw \citep{openclaw}, Codex CLI \citep{codexcli}, Aider \citep{aider}, OpenHands \citep{openhands}, Cline \citep{cline}, Crush \citep{crush}, and Cursor \citep{cursor}) and extract candidate features from it. The second phase examines each candidate feature from the first phase in greater detail, filling in invocation form, parameter semantics, and boundary behavior. The outputs of the two phases are merged, deduplicated, and pruned into one inventory of frontier-harness capabilities.

\looseness=-1 %
\dthrow{Turning the inventory into clauses.} As shown in Table~\ref{tab:doc_chapters}, the development document covers the deliverable and runtime interface, built-in tools, top-level commands and subcommands, permission and role modes, global flags, in-session commands and shortcuts, persistent memory, convention files, consistency rules, exit codes, and the API for driving automated evaluation. Each clause is given a syntax, a meaning, and an expected behavior. Confining each clause in this way serves two purposes, (1) verifiability from external behavior and (2) independence from implementation. To make the first purpose operational, we require the effect of every feature to be observable through one of four channels, namely the requests sent to the model, the files on disk, the process exit code, and the command output. To serve the second purpose, apart from the three contracts of the contract skeleton we do not constrain the manner of implementation.

\begin{table}[t]
\centering
\small
\renewcommand{\arraystretch}{1.15}
\caption{Chapter structure of the development document.}
\label{tab:doc_chapters}
{\setlength{\tabcolsep}{0pt}%
\begin{tabularx}{\textwidth}{@{}X>{\hspace{12pt}}X@{}}
\toprule
\rowcolor{d2hTint} %
\textbf{Content} & \textbf{Principal clause types} \\
\midrule
Deliverable and runtime interface & File structure, entry script \\
Built-in tools & Invocation form and boundary behavior of tools \\
Top-level commands and subcommands & Command syntax and exit codes \\
Permission modes, role modes, sandbox and lists & Five permission modes, five role modes, deny rules \\
Global flags & Semantics of flags, and their effect on requests \\
In-session commands and shortcuts & Command dispatch and terminal interaction \\
Persistent memory & Writing, retrieval, and clearing of memory \\
Convention files & Discovery order and injection method of files \\
Consistency rules & Relational constraints across chapters \\
Exit codes & A closed set of values \\
\bottomrule
\end{tabularx}}
\end{table}

\dthrow{Disambiguating features.} The clauses from the previous step come from twelve sources, and to avoid semantic conflicts we use the following two disambiguation rules.
(1) We merge commands with overlapping semantics. For example, Claude Code names context compaction \texttt{/compact} whereas Cursor names it \texttt{/summarize}, and the development document adopts \texttt{/compact}.
(2) We split distinct concepts that occupy the same name. Take the word \textit{plan}: in Claude Code it is one of six permission modes and means read-only, whereas in Aider, proposing changes and carrying out edits are handled by two model roles, architect and editor, and Aider's documentation instead specifies the responsibilities assigned to each role. The development document accordingly assigns the read-only constraint to permissions and responsibility for proposing and applying changes to roles, so that permissions and roles each form their own clause.

\subsubsection{Contract Skeleton}
\label{sec:skeleton}

The form of the contract skeleton determines what proportion of the candidate harness the build model writes. To maximize this proportion while keeping the candidate harness observable, we adopt a stripped-down mini-swe-agent \citep{minisweagent} as the skeleton.

\looseness=-1 %
\dthrow{Composition.} The contract skeleton consists of five Python source files (roughly seventy lines in total), plus one install script, one entry script, and one project config. Three of its interfaces constitute an immutable external contract, namely (1) the CLI, (2) the headless entry point, and (3) the API for driving automated evaluation. A representative candidate harness runs to approximately 30 Python source files and approximately 16,000 lines, of which roughly 99.6\% is written by the build model beyond what the contract skeleton supplies.

\looseness=-1 %
\dthrow{Construction environment.} Every build takes place in a newly created empty container, into which the contract skeleton and the development document are copied and which has no external network access other than the designated model gateway. Before construction begins we run a preflight check confirming that the container holds no unexpected files and that OpenCode starts normally; failure of either condition aborts the build and produces no result.

\subsection{Test Case Synthesis}
\label{sec:synthesis}

\begin{figure}[t]
\centering
\includegraphics[width=\textwidth]{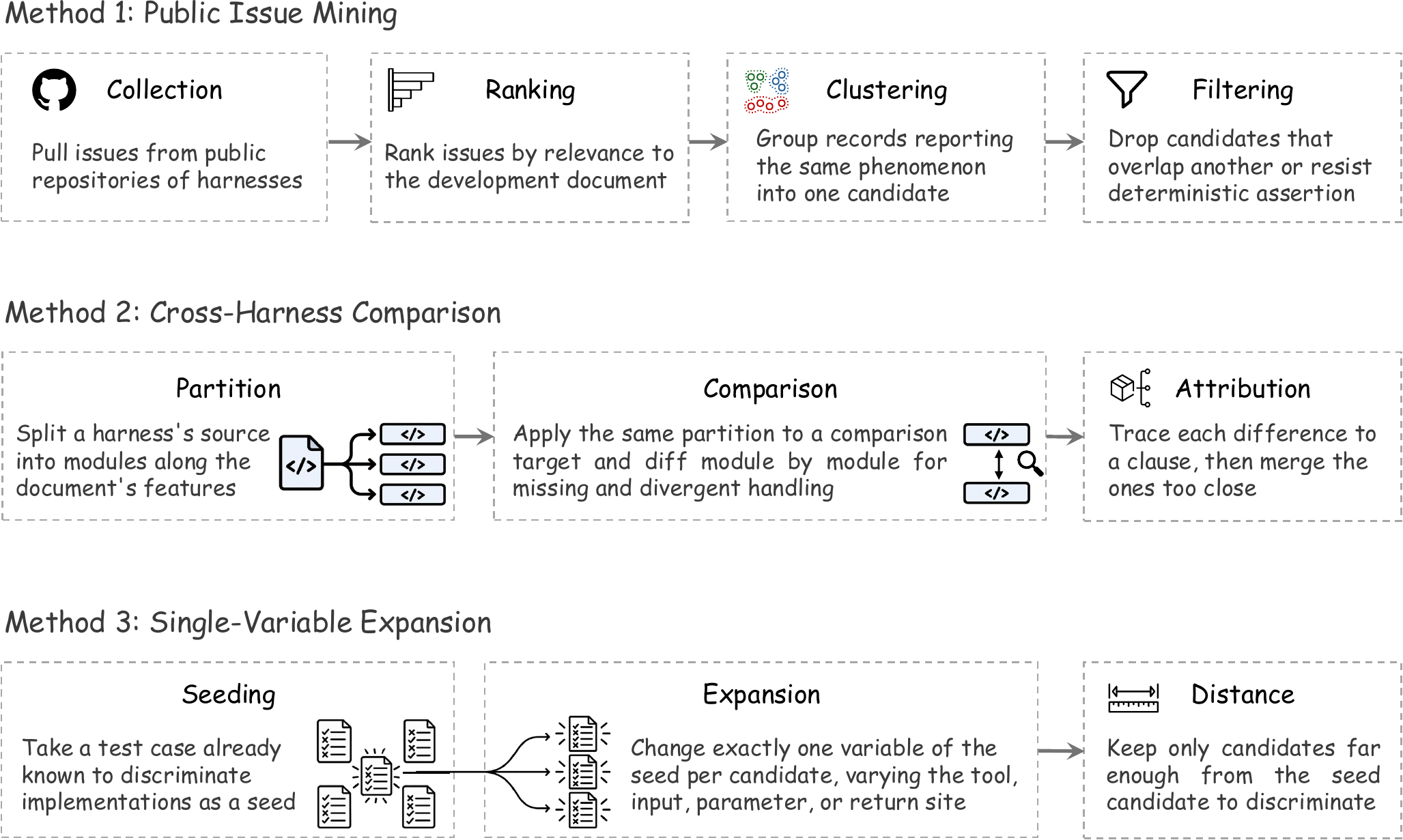}
\caption{The three test case synthesis methods.}
\label{fig:synthesis}
\end{figure}

\looseness=-1 %
At design time, a test case must satisfy two requirements at once: it must be anchored to some clause of the development document, and it must discriminate between implementations; whether a retained case actually discriminates is settled only after evaluation. As shown in Figure~\ref{fig:synthesis}, we use three methods in a yield-driven cycle, the first two broadening the coverage of the test cases and the third deepening what they inspect. \S\ref{sec:mining} draws on the public records of harness users, \S\ref{sec:comparison} on the source of mature harnesses, and \S\ref{sec:expansion} on the existing test suite itself. We move to the next method when the current method's marginal yield declines. Because the methods draw on different sources, they also use different filtering criteria.

\subsubsection{Public Issue Mining}
\label{sec:mining}

This method rests on an assumption about the distribution of defects. LLMs are trained on code written by human developers. Mature harnesses and the harnesses written by build models face the same interface semantics and the same class of boundary conditions. We therefore expect substantial overlap between the defects mature harnesses have exposed over years of evolution and the mistakes made by the harnesses written by build models, which makes public issue records a ready source of candidates.

\dthrow{Collection.} We pull every issue from the public repositories of four frontier harnesses (Claude Code, OpenCode, Gemini CLI, and Kimi Code); after filtering out pull requests these number 9{,}868, 5{,}438, 4{,}831, and 946 respectively, 21{,}083 in total.

\dthrow{Ranking and stopping.} More than 20{,}000 records cannot be reviewed manually in full, so we first rank them by relevance to the development document. Manual review starts at the top of the ranking and stops when the marginal yield falls below the cost of further review.

\looseness=-1 %
\dthrow{Clustering.} The same defect is often reported repeatedly, worded differently yet all pointing at one feature. We cluster the records we have read by the phenomenon they report, forming one candidate test case per cluster, into whose material the records of that cluster are then merged.

\dthrow{Filtering.} Distinct phenomena may still land on the same point of the implementation. Where two clusters anchor to the same clause and the effects they observe are too close, the test cases realized from them would pass and fail together on the same set of implementations, so only the one with more complete material is kept. A remaining candidate is retained only if it can additionally be turned into a deterministic assertion and is not yet covered by the existing test suite.

This method's net yield is high at the outset and falls as the same sources are mined further, on which basis we turn to the next method.

\subsubsection{Cross-Harness Comparison}
\label{sec:comparison}

The previous method is limited by the quantity and quality of external records; this method broadens coverage from the source side instead. We compare one candidate harness against the public source of a mature harness and examine which capabilities that candidate harness lacks. We call the mature implementation that serves as the reference the comparison target; examples include Kimi Code and Qwen Code.

\dthrow{Partition.} A candidate harness's source is partitioned into modules according to the features the development document lists, each module corresponding to one group of features in the document.

\looseness=-1 %
\dthrow{Comparison.} The same partition is applied to the source of the comparison target, which places the two on a common set of functional boundaries, and we enumerate their implementation differences module by module, covering both the features the comparison target handles and the candidate harness does not, and the boundaries on which the two behave differently.

\looseness=-1 %
\dthrow{Attribution.} A difference does not necessarily stem from a defect; it may merely reflect a different manner of implementation. Every difference is traced back to the development document, and only those derivable from the purpose of some clause count as candidate test cases.

\looseness=-1
The coverage of this method is capped by the functional surface of the comparison target: once the candidate test cases obtainable from comparison are exhausted, we turn to the next method.

\subsubsection{Single-Variable Expansion}
\label{sec:expansion}

This method starts from a test case already known to discriminate and changes exactly one thing each time, so as to deepen what the test cases inspect.

\dthrow{Seed selection.} We take a test case already shown to discriminate as the seed test case; the defect it exposes is the target of that expansion round.

\dthrow{Expansion.} A test case consists of a scenario and an assertion; the scenario is the session through which the mock model drives the candidate harness, and it fixes only the setting and the flow, while the adjudication rests with the assertion. We vary only the scenario, keeping the assertion fixed. Each time we change exactly one variable of the seed scenario to obtain a new candidate, and every candidate must still be anchored to some clause of the development document. We use four variables.

\begin{itemize}[leftmargin=2em]
\item \textit{The tool.} A coding agent can often reach the same end through several tools, while a candidate harness may implement only some of them correctly.
\item \textit{The input form.} The same inspection can be triggered by inputs of several forms; when inspecting path escape, we can replace the input with a symlink to a file, a symlink to a directory, a multi-level symlink, or a relative-path symlink.
\item \textit{The erroneous parameter.} The same coding mistake may appear on different parameters, for instance treating the value 0 as absent.
\item \textit{The site where the result is returned.} The result of the same behavior may appear on standard output, on standard error, or in the error text.
\end{itemize}

\dthrow{Distance check.} Not every modified scenario adds discriminative power beyond the seed, so we assess whether each candidate's scenario is far enough from the seed scenario. A candidate too close to the seed is discarded; otherwise we retain it as a test case and use it as the next seed, from which the following round of single-variable expansion proceeds.

This method spreads outward from seeds, complementing the two preceding methods. A clause with no anchored test case cannot provide a seed, and such gaps can only be filled by the first two methods; the three are therefore used in alternation.

\subsection{Positive Control}
\label{sec:pc}

\looseness=-1 %
A test case that passes does not necessarily show that the candidate harness has implemented the requirement in question. Consider path escape: the mock model's script has the candidate harness read a symlink pointing outside the workspace with its read tool, and the test case asserts that the external file's content does not appear in the requests sent to the model. That the content is absent may mean the implementation blocked this out-of-bounds read, or it may mean its read tool cannot read any file at all. The assertion holds in both situations; only the former constitutes meeting the requirement, and the latter is a vacuous pass.

To rule out vacuous passes, we design the positive control mechanism. A positive control shares the scenario and the fixture with its test case, where the fixture consists of the files and initial state prepared in the workspace before a run. It changes only the input or access route, selects a case outside the constraint imposed by the clause under test, and asserts an outcome that must occur when the observation precondition holds, that is, when the effect the test case inspects is in fact observable. In the example above, the positive control reuses the same symlink, accesses it through the bash tool instead, and asserts that the external file is indeed readable.

We group test cases that share an observation precondition under one positive control, and a test case with several such preconditions is gated by several controls. If a positive control fails, the precondition does not hold, so a pass by the test cases in its group can only be vacuous and they are recorded as FAIL regardless of their own statuses. Positive controls themselves do not count toward the pass rate.

Given that the assertion is correctly constructed, a vacuous pass has three causes, each ruled out by one class of positive control.

\looseness=-1 %
\textit{Channel.} The tool's result may never reach the channel being observed at all. A test case observes only one of the four channels; if an implementation's read tool never places file content into a request, then the content of a file outside the workspace will not appear there either. We therefore read an ordinary file with the read tool and confirm that its content does appear in the requests sent to the model. Failure here means the implementation does not put tool results onto this channel at all, so the absence of the forbidden content does not constitute interception. The path-escape test case does not require a control of its own, because its reverse control, described below, already covers this condition.

\textit{Same form.} This rules out failures that would have occurred even without the constraint under test. Reusing the same input form and changing only the route of access, we access the same symlink through the implementation's bash tool and confirm that it resolves outside the workspace and that the file there is readable (the path-escape clause constrains file tools, and commands executed by bash fall outside its scope). Failure here means the implementation cannot complete even an access unconstrained by the clause, so its miss on the access under test has nothing to do with interception.

\textit{Reverse.} This rules out the case where the implementation refuses inputs of this kind categorically, on grounds other than this access being out of bounds. An implementation that refuses all symlinks passes the test case above, yet it also refuses symlinks pointing inside the workspace, which are valid in normal use. We therefore replace the input with such a symlink, perform the same read, and confirm that its content still appears in the requests sent to the model. Failure here means the pass did not arise from recognizing an out-of-bounds read.

\subsection{Statistics}
\label{sec:statistics}

\begin{figure}[t]
\centering
\begin{subfigure}[t]{0.475\textwidth}
\centering
\includegraphics[width=\textwidth]{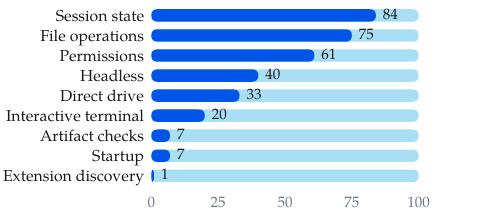}
\caption{Test cases per category.}
\label{fig:suite:a}
\end{subfigure}
\hfill
\begin{subfigure}[t]{0.475\textwidth}
\centering
\includegraphics[width=\textwidth]{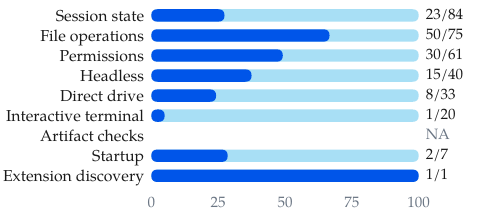}
\caption{The share of test cases gated by at least one positive control.}
\label{fig:suite:b}
\end{subfigure}
\caption{File operations and permissions have the highest positive-control coverage, whereas the interactive terminal has the lowest. Artifact checks is marked NA, because the contract check already settles in full what its test cases examine, namely that the deliverable is complete.}
\label{fig:suite}
\end{figure}

The test suite contains 328 test cases, each accompanied where necessary by positive controls. We partition them into nine mutually disjoint categories according to the content of the development document and the three external contracts of the contract skeleton, as shown in Figure~\ref{fig:suite}.

\dthrow{Coverage of the positive controls.} Of the 328 test cases, 130 are gated by at least one positive control (Figure~\ref{fig:suite}b), of which 114 are gated by one, 15 by two, and one by three. The control with the widest scope gates six test cases and confirms that a symlink pointing outside the workspace is in fact readable; the next two gate five each and confirm, respectively, that reading by range works and that a search returns a hit. These three thus check the observation preconditions shared by the most test cases.

\section{Experimental Setup}
\label{sec:setup}

\dthrow{Models under evaluation.} We evaluate nineteen models drawn from eleven vendors, namely Claude Opus 4.6 \citep{opus46}, Opus 4.7 \citep{opus47}, Opus 4.8 \citep{opus48}, and Opus 5 \citep{opus5}; GPT-5.5 \citep{gpt55}, GPT-5.6-sol \citep{gpt56sol}, and GPT-6 Astra \citep{gpt6astra}; Qwen3.8 Max 0902 \citep{qwen38max}; Kimi K3 \citep{kimik3}; GLM 5.2 \citep{glm52} and GLM 5.3 \citep{glm53}; DeepSeek V4 Pro \citep{deepseekv4pro} and DeepSeek V4.1 Flash \citep{deepseekv41flash}; Hy3 \citep{hy3} and Hy4 preview \citep{hy4preview}; Doubao Seed 2.1 Pro \citep{doubaoseed21pro}; MiMo-V2.6-Flash \citep{mimo26flash}; Grok 4.7 \citep{grok47}; and Step 5 Preview \citep{step5preview}.

\dthrow{Execution environment.} Both stages run on the same server, an AMD EPYC 7K83 machine with 48 logical processors and 128 GB of memory.

\dthrow{Configuration.} In Stage~1, the builder harness is OpenCode 1.17.0, installed into the build container as a single compiled executable. The builder constructs the candidate harness through OpenCode's non-interactive entry point \texttt{opencode run}, with the task description supplied once on the command line. Because that entry point advances the session by only one turn per invocation, we resume the same session with \texttt{-{}-continue}, as OpenCode's official documentation describes, and supply a further prompt whenever no submission has yet been made, up to twenty-four turns or until the wall clock is exhausted. As the session approaches the context limit, the builder harness compacts the history automatically, with the trigger threshold set at eighty percent of the model's context window. Each build is subject to a two-hour wall-clock limit and to a quota of at most three submissions, and terminates as soon as either is exhausted. In Stage~2, every test case is executed once inside its own container, on its own fixture, subject to a 900-second limit beyond which execution is terminated and the test case recorded as FAIL.

\section{Results}
\label{sec:results}

\subsection{Main Results}
\label{sec:main}

\begin{table}[t]
\centering
\small
\renewcommand{\arraystretch}{1.12}
\caption{Pass rate of the nineteen models (\%), decomposed by evaluation category. The rightmost column is the primary measure: the pass rate over all 328 test cases, reported as the mean $\pm$ the standard deviation over the three scored candidate harnesses. ``Other'' merges artifact checks, startup, and extension discovery, which hold 7, 7, and 1 test cases respectively. Positive controls are excluded from every denominator. Abbreviations in the header are Perm.\ for permissions, File ops.\ for file operations, Session for session state, Direct for direct drive, and Interactive for interactive terminal.}
\label{tab:main_results}
\setlength{\tabcolsep}{0pt}\footnotesize
\begin{tabularx}{\textwidth}{@{}X>{\hspace{4.4pt}}c>{\hspace{4.4pt}}c>{\hspace{4.4pt}}c>{\hspace{4.4pt}}c>{\hspace{4.4pt}}c>{\hspace{4.4pt}}c>{\hspace{4.4pt}}c>{\hspace{4.4pt}}c@{}}
\toprule
\multirow{2}{*}{\textbf{Model}} & \cellcolor{d2hTint}\textbf{Perm.} & \cellcolor{d2hTint}\textbf{File ops.} & \cellcolor{d2hTint}\textbf{Session} & \cellcolor{d2hTint}\textbf{Direct} & \cellcolor{d2hTint}\textbf{Headless} & \cellcolor{d2hTint}\textbf{Interactive} & \cellcolor{d2hTint}\textbf{Other} & \cellcolor{d2hTint}\textbf{All} \\
\cmidrule(lr){2-9}
 & \textbf{61} & \textbf{75} & \textbf{84} & \textbf{33} & \textbf{40} & \textbf{20} & \textbf{15} & \textbf{328} \\
\midrule
GPT-6 Astra & \textbf{71.6} & 46.7 & \textbf{44.0} & \textbf{39.4} & \textbf{45.0} & 91.7 & 60.0 & \textbf{53.05 $\pm$ 0.25} \\
Opus 5 & 63.4 & 39.6 & 38.1 & 17.2 & 43.3 & 91.7 & 75.6 & \textbf{46.65 $\pm$ 1.80} \\
Qwen3.8 Max & 61.2 & 38.2 & 43.3 & 15.2 & 32.5 & \textbf{93.3} & 75.6 & \textbf{45.83 $\pm$ 0.14} \\
GPT-5.6-sol & 47.5 & 48.9 & 29.4 & 32.3 & 39.2 & 83.3 & 68.9 & \textbf{43.80 $\pm$ 2.44} \\
Grok 4.7 & 54.6 & \textbf{50.7} & 28.2 & 22.2 & 33.3 & 78.3 & 75.6 & \textbf{43.50 $\pm$ 2.49} \\
GLM 5.3 & 51.4 & 41.8 & 33.3 & 18.2 & 31.7 & 85.0 & 77.8 & \textbf{42.07 $\pm$ 1.39} \\
MiMo-V2.6-Flash & 50.3 & 40.4 & 37.7 & 14.1 & 33.3 & 76.7 & 75.6 & \textbf{41.87 $\pm$ 3.29} \\
Step 5 Preview & 47.0 & 45.8 & 34.1 & 14.1 & 25.8 & 76.7 & 77.8 & \textbf{40.75 $\pm$ 3.29} \\
DeepSeek V4.1 Flash & 49.2 & 36.4 & 34.5 & 14.1 & 31.7 & 81.7 & 80.0 & \textbf{40.24 $\pm$ 0.86} \\
Hy4 preview & 44.8 & 38.2 & 35.7 & 10.1 & 27.5 & 78.3 & 77.8 & \textbf{38.92 $\pm$ 1.46} \\
Kimi K3 & 43.2 & 40.9 & 25.4 & 18.2 & 32.5 & 86.7 & 73.3 & \textbf{38.31 $\pm$ 1.28} \\
Opus 4.8 & 30.1 & 48.4 & 33.7 & 12.1 & 22.5 & 83.3 & 80.0 & \textbf{38.01 $\pm$ 0.87} \\
GPT-5.5 & 36.1 & 36.4 & 25.0 & 24.2 & 20.8 & 76.7 & 77.8 & \textbf{34.65 $\pm$ 1.42} \\
Doubao Seed 2.1 Pro & 41.5 & 36.0 & 15.5 & 20.2 & 18.3 & 83.3 & 73.3 & \textbf{32.62 $\pm$ 1.51} \\
Opus 4.7 & 36.1 & 29.8 & 25.8 & 12.1 & 16.7 & 85.0 & 80.0 & \textbf{32.22 $\pm$ 1.66} \\
DeepSeek V4 Pro & 29.5 & 33.3 & 21.0 & 23.2 & 21.7 & 78.3 & 80.0 & \textbf{31.91 $\pm$ 1.23} \\
Opus 4.6 & 20.8 & 37.3 & 21.8 & 25.3 & 3.3 & 60.0 & 80.0 & \textbf{28.25 $\pm$ 0.14} \\
GLM 5.2 & 23.0 & 32.9 & 21.8 & 12.1 & 12.5 & 63.3 & 80.0 & \textbf{27.64 $\pm$ 1.01} \\
Hy3 & 28.4 & 31.1 & 19.4 & 13.1 & 9.2 & 65.0 & 80.0 & \textbf{27.44 $\pm$ 0.43} \\
\midrule
\textbf{Mean} & \textbf{43.7} & \textbf{39.6} & \textbf{29.9} & \textbf{18.8} & \textbf{26.4} & \textbf{79.9} & \textbf{76.3} & \textbf{38.30} \\
\bottomrule
\end{tabularx}
\end{table}

\looseness=-1 %
\dthrow{The suite is far from saturated.} As shown in Table~\ref{tab:main_results}, GPT-6 Astra attains the highest pass rate at 53.05\%, while the weakest model, Hy3, reaches 27.44\%, a gap of roughly twenty-six percentage points. Even the strongest model fails nearly half of the test cases. The suite is therefore neither saturated on this cohort nor incapable of discriminating among the models.

\dthrow{The top two models differ on boundary scenarios.} On the ordinary scenarios Opus 5 leads with 80.68\% against 72.46\% for GPT-6 Astra, whereas on the boundary scenarios GPT-6 Astra leads with 48.97\% and Opus 5 falls to 37.02\%. The performance gap between the two models is therefore concentrated in their handling of boundary clauses.

\subsection{Results by Category}
\label{sec:bylayer}

In addition to each category's average pass rate, we report the gap between its strongest and weakest models, which we call the strong-weak gap.

\begin{table}[t]
\centering
\small
\renewcommand{\arraystretch}{1.12}
\caption{Strong-weak gap and test-case composition by evaluation category. The last three columns split each category's test cases into floor test cases, which no candidate harness passes, ceiling test cases, which every candidate harness passes, and the remaining discriminating test cases that determine the ranking.}
\label{tab:by_category}
{\setlength{\tabcolsep}{0pt}%
\begin{tabularx}{\textwidth}{@{}X>{\hspace{12pt}}c>{\hspace{12pt}}c>{\hspace{12pt}}c>{\hspace{12pt}}c>{\hspace{12pt}}c>{\hspace{12pt}}c@{}}
\toprule
\rowcolor{d2hTint} %
\textbf{Category} & \textbf{Cases} & \textbf{Avg.\ pass} & \textbf{Strong-weak} & \textbf{Floor} & \textbf{Ceiling} & \textbf{Discriminating} \\
\midrule
Session state & 84 & 29.9 & 28.6 & 14 & 0 & 70 \\
File operations & 75 & 39.6 & 20.9 & 3 & 0 & 72 \\
Permissions & 61 & 43.7 & 50.8 & 10 & 0 & 51 \\
Headless & 40 & 26.4 & 41.7 & 9 & 0 & 31 \\
Direct drive & 33 & 18.8 & 29.3 & 17 & 0 & 16 \\
Interactive terminal & 20 & 79.9 & 33.3 & 0 & 7 & 13 \\
Artifact checks & 7 & 100.0 & 0.0 & 0 & 7 & 0 \\
Startup & 7 & 63.4 & 42.9 & 2 & 2 & 3 \\
Extension discovery & 1 & 0.0 & 0.0 & 1 & 0 & 0 \\
\midrule
\textbf{Total} & \textbf{328} & \textbf{38.30} & & \textbf{56} & \textbf{16} & \textbf{256} \\
\bottomrule
\end{tabularx}}
\end{table}

\dthrow{The categories differ sharply in discriminative power.} As shown in Table~\ref{tab:by_category}, the permissions category provides the greatest separation among models: its average pass rate is 43.7\%, its strong-weak gap is 50.8 points, and 51 of its 61 test cases are discriminating. File operations has a comparable average pass rate of 39.6\% but a gap of only 20.9 points. The interactive-terminal category averages 79.9\% and is approaching saturation. In addition, 56 test cases are failed by every candidate harness, so these cases provide no discrimination within the evaluated cohort; they are concentrated in the direct drive, session state, permissions, and headless categories.

\dthrow{A model's rank varies greatly across categories.} Among the six main categories, each of which holds at least twenty test cases, a model's rank varies by as many as sixteen positions: Grok 4.7 ranks first on file operations but eleventh on session state, Opus 4.6 third on direct drive but nineteenth on permissions, and Hy4 preview fifth on session state but nineteenth on direct drive. GPT-6 Astra leads four of the six categories, while Grok 4.7 and Qwen3.8 Max lead the remaining two. \textbf{The categories show positive pairwise associations rather than clear trade-offs.} All fifteen pairwise correlations between the six main categories are positive, with a mean of $+0.521$, and the overall pass rate correlates with all six of them, five between 0.69 and 0.94. These results are consistent with a largely one-dimensional capability: models differ mainly in overall capability rather than in which categories they specialize in. Model rankings nevertheless vary substantially across categories.

\dthrow{The test suite is a reliable instrument.} The KR-20 coefficient computed over the fifty-seven candidate harnesses across all test cases is 0.913, indicating high internal consistency, as expected if the test cases measure one underlying capability.

\subsection{Architecture of the Candidate Harnesses}
\label{sec:arch}

This section analyzes the source code of the fifty-seven scored candidate harnesses. A module is a Python source file under the root of a candidate harness and a code line is a non-blank line; we define a category as having a dedicated module when at least one module filename contains a keyword associated with that category.

\begin{table}[t]
\centering
\small
\renewcommand{\arraystretch}{1.12}
\caption{Architecture of the candidate harnesses. Every column is a mean over each model's three scored builds, except \emph{Lines/module}, which is mean code lines divided by mean module count, computed before rounding; \emph{Largest module} is the share of all code lines held by the single largest module.}
\label{tab:architecture}
{\setlength{\tabcolsep}{0pt}%
\begin{tabularx}{\textwidth}{@{}X>{\hspace{12pt}}c>{\hspace{12pt}}c>{\hspace{12pt}}c>{\hspace{12pt}}c>{\hspace{12pt}}c@{}}
\toprule
\rowcolor{d2hTint} %
\textbf{Model} & \textbf{Pass rate} & \textbf{Code lines} & \textbf{Modules} & \textbf{Lines/module} & \textbf{Largest module} \\
\midrule
GPT-6 Astra & 53.05 & 7964 & 12.3 & \textbf{646} & 21\% \\
Opus 5 & 46.65 & 13845 & 33.3 & 415 & 13\% \\
Qwen3.8 Max & 45.83 & \textbf{16681} & 40.0 & 417 & 10\% \\
GPT-5.6-sol & 43.80 & 2745 & 14.7 & 187 & 23\% \\
Grok 4.7 & 43.50 & 5625 & 13.3 & 422 & 25\% \\
GLM 5.3 & 42.07 & 8186 & 21.3 & 384 & 16\% \\
MiMo-V2.6-Flash & 41.87 & 11063 & 23.0 & 481 & 18\% \\
Step 5 Preview & 40.75 & 7620 & 18.7 & 408 & 16\% \\
DeepSeek V4.1 Flash & 40.24 & 5519 & 15.3 & 360 & 20\% \\
Hy4 preview & 38.92 & 7251 & 19.7 & 369 & 16\% \\
Kimi K3 & 38.31 & 5967 & 23.7 & 252 & 14\% \\
Opus 4.8 & 38.01 & 5881 & 30.3 & 194 & 13\% \\
GPT-5.5 & 34.65 & \textbf{1271} & \textbf{2.7} & 477 & 78\% \\
Doubao Seed 2.1 Pro & 32.62 & 6761 & \textbf{41.7} & \textbf{162} & 8\% \\
Opus 4.7 & 32.22 & 4355 & 18.7 & 233 & 19\% \\
DeepSeek V4 Pro & 31.91 & 4169 & 13.0 & 321 & 27\% \\
Opus 4.6 & 28.25 & 2351 & 12.0 & 196 & 24\% \\
GLM 5.2 & 27.64 & 4047 & 18.3 & 221 & 25\% \\
Hy3 & 27.44 & 2690 & 13.7 & 197 & 26\% \\
\bottomrule
\end{tabularx}}
\end{table}

\begin{figure}[t]
\centering
\includegraphics[width=\textwidth]{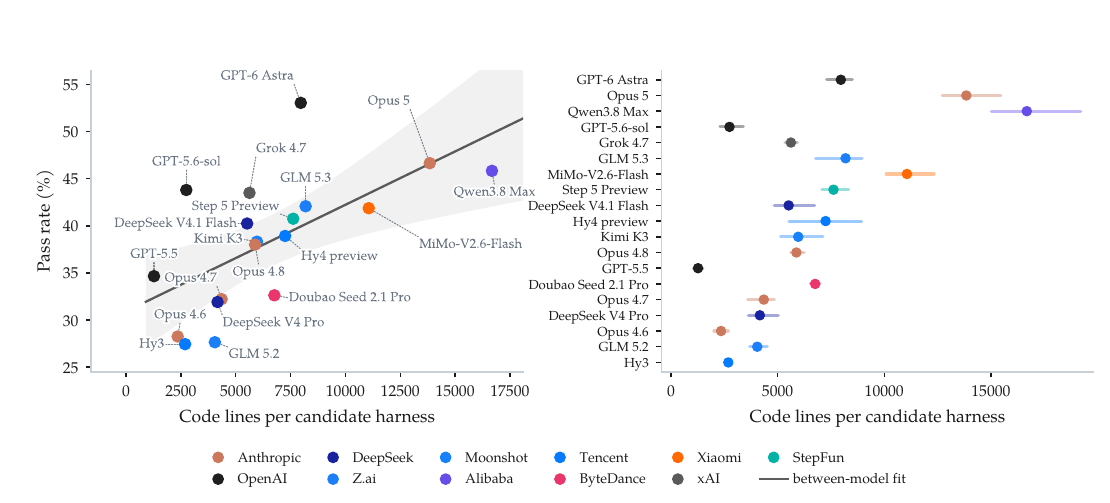}
\caption{Code lines against pass rate. Left, one point per model at its mean over the three scored builds, with a least-squares fit and its 95\% confidence band. Right, each model's three scored builds, spanned by a segment with the mean marked, one row per model, ordered by pass rate.}
\label{fig:size}
\end{figure}

\dthrow{Implementation scale is set mainly by the model.} As shown in Table~\ref{tab:architecture}, per-model mean code size ranges from 1{,}271 to 16{,}681 lines, and across the fifty-seven builds the range is 1{,}062 to 19{,}208. Most of this variation lies between models rather than among the three builds of a given model: 95.9\% of the variance in code lines is between models, and the coefficient of variation across the three builds of a model averages 12.7\%, compared with 60.1\% across the nineteen model means. The corresponding between-model variance shares are 87.9\% for module count, 80.6\% for lines per module, and 95.7\% for largest-module share; Figure~\ref{fig:size} shows the same between-model dominance for code lines. The correlation between code lines and pass rate is 0.614 across the fifty-seven builds, and 0.629 between the model means. Implementation scale is driven mainly by model identity, with comparatively little variation across a model's three builds.

\dthrow{Harnesses with more modules score lower.} Controlling for code lines, the partial correlation between module count and pass rate is $-0.423$, and the leave-one-model-out check keeps it between $-0.484$ and $-0.309$; correspondingly, lines per module has a partial correlation of $+0.504$, holding between 0.312 and 0.649 under the same check. Fragmentation is therefore not free.

\section{Related Work}
\label{sec:relatedwork}

\subsection{Code and Software Engineering Benchmarks}
\label{sec:rw-swe}

Coding benchmarks have evolved toward tasks involving increasingly large codebases, and fall broadly into two classes: code completion and repository-level generation.

Benchmarks in the completion class ask for local edits to existing code. HumanEval \citep{humaneval} established the practice of judging a single function by unit tests, and MBPP \citep{mbpp} and EvalPlus \citep{evalplus} extended this unit-test-based paradigm to a larger scale; the scope then grew to a whole repository with SWE-bench \citep{swebench}, which asks the model to resolve a real issue and judges by the tests shipped with the corresponding pull request. Variants of SWE-bench extend this setup while addressing contamination \citep{swebenchpro}, evaluating end-to-end performance on freelance tasks \citep{swelancer}, or deriving specifications from documentation changes \citep{doc2feat}.

Repository-level generation removes the existing implementation and asks for a complete one. The evaluation criterion depends on the available reference implementation. Commit0 \citep{commit0} and NL2Repo-Bench \citep{nl2repobench} ask models to reconstruct released libraries and judge by the supplied or upstream unit tests. RepoZero \citep{repozero} compares output equivalence against a second implementation; ProgramBench \citep{programbench} uses a reference executable to generate end-to-end tests through fuzzing; and DeepSWE \citep{deepswe} hand-writes a verifier per task that accepts any implementation of the requested feature set.

\subsection{Harness Benchmarks}
\label{sec:rw-harness}

A harness is a class of software whose functional surface can be enumerated feature by feature \citep{whatmakesharness,harnesssurvey,harnesseng}. This makes it a suitable target for a construction benchmark. Most harness benchmarks evaluate existing harnesses. Claw-SWE-Bench \citep{clawswebench} fixes the prompt, budget, workspace, and evaluator, and finds that the choice of harness shifts scores by 27.4 points, compared with 29.4 points for the choice of model; Harness-Bench \citep{harnessbench} isolates configuration-level effects over 106 sandboxed tasks and concludes that ability should be reported at the model--harness configuration level rather than attributed to the base model alone.

Other benchmarks hold the harness fixed and evaluate a single capability of it, such as rule compliance across the surfaces an agent reads \citep{harnessif}, skill retrieval under progressive disclosure \citep{skilluse}, component-lifecycle reasoning \citep{cordisbench}, or safety along a policy ladder and across persistent carriers \citep{permissiondenied,harnessrisk,harnesssafe}. Still others let the given harness change from stage to stage and ask whether the agent keeps pace \citep{evoharnessbench}.

A further line of work asks a model to produce or revise the harness itself, starting from an existing harness and its execution feedback \citep{harnessdev,evobench,harnessopt}; because gains are measured on downstream tasks, they cannot be separated from those obtained by simply allowing more test-time search, a confound noted in this literature \citep{harnessevaleval}. All of these studies infer harness quality from downstream tasks. Such tasks must be prepared in advance, and the ability of the model that drives them also influences the measured score. Nor is output-only scoring sufficient, since a harness can return a correct and harmless answer along a trajectory that reached unauthorized resources, and output-level evaluation cannot see failures of that kind \citep{harnesssafetyaudit}. None of them adjudicates, clause by clause, whether a harness's own behavior conforms to a specification.

\subsection{Position of This Work}
\label{sec:rw-position}

Doc2Harness differs from the work above in three respects. First, the target deliverable does not correspond to any existing implementation: the development document consolidates the functionality of twelve frontier harnesses into a self-consistent set of clauses, whereas prior repository-level deliverables are reconstructions of something already published. Second, the candidate harness is driven by a mock model following a preset script rather than by a real model on downstream tasks, so the evaluation is deterministic and what is judged is the candidate harness itself. Third, the candidate harness is executed end to end as a black box through a complete scripted session, with effects read from four channels. This makes constraints of the form ``a rejected operation must not take place'' adjudicable clause by clause; and because an assertion requiring some effect not to appear may hold simply because the feature under test never ran, such test cases are equipped with positive controls confirming that the observation precondition holds. In sum, Doc2Harness is the first benchmark able to assess deterministically how well a model develops a harness.

\section{Limitations}
\label{sec:limitations}

\dthrow{The protocol measures a single build.} Each build allows up to three submissions and is not an iterative build--evaluate--repair loop. The contract check tests form rather than behavior, so the amount of revision a candidate harness receives depends on when the model submits, and the pass rate of an early submitter is a lower bound rather than an upper bound. The protocol does not measure improvement based on behavioral feedback after the candidate harness is frozen.

\dthrow{A single defect can fail several test cases at once.} Some test cases are built around the same feature: the single-variable expansion of \S\ref{sec:expansion} derives several variants from one seed test case, and although the distance check discards candidates too close to their seed, the variants that remain are still correlated. A single defect can therefore cause correlated failures across these variants.

\section{Conclusion}
\label{sec:conclusion}

Doc2Harness is the first benchmark able to assess deterministically how well a model develops a harness. The build model under evaluation, operating inside a fixed builder harness, builds a complete candidate harness from a development document and a contract skeleton that declares interfaces only; a mock model then drives it through a preset script for each test case, and explicit assertions determine behavioral conformance.

Our experiments show that even the strongest model fails nearly half of the test cases, and that the failures are not random but concentrate on one implementation choice: whether the same functionality is split across several modules. We hope Doc2Harness will serve as a benchmark for measuring this class of construction ability.

\bibliographystyle{colm2026_conference}
\bibliography{ref}

\end{document}